\documentclass[a4paper,12pt]{article}
\newtheorem{theorem}{Theorem}[section]

\newtheorem{remark}{Remark}[section]  
\newtheorem{assumption}{Assumption}[section]  

\title{On the Constitutive Relations for Second Sound in Thermo-Electroelasticity}

\author{A. Montanaro \\ montanaro$@$dmsa.unipd.it
\\University of Padua, Italy}

\bigskip

\begin{document}  

\maketitle



\maketitle

\centerline{\bf{Abstract}}
Extending previous papers of Coleman-Fabrizio-Owen  \cite{RefColFabOwen1}, \cite{RefColFabOwen2}  and \"{O}nc\"{u}-Moodie \cite{RefOncuMoodie}, we give a derivation of the thermodynamic restrictions on the constitutive relations of an electrically polarizable and finitely deformable heat conducting elastic continuum, interacting with the electric field. 
The constitutive equations include an evolution equation for the heat flux; the latter and the temperature obey a frame-invariant form of Cattaneo's equation.

\section{Introduction}
 
In some papers Coleman, Fabrizio and Owen \cite{RefColFabOwen1}, \cite{RefColFabOwen2} gave a derivation of the implications of the second law of thermodynamics for a rigid heat conductor for which the heat flux vector ${\bf q}$ and the temperature $\theta$ obey the  relation 
\begin{equation}   \label{eq:CFOrel}
	\hat{\bf T}(\theta) \dot {\bf q} + {\bf q} = - \hat{\bf K}(\theta) grad \theta \, ,
\end{equation}
 with $\hat{\bf T}(\theta)$ and $\hat{\bf K}(\theta)$ non-singular, that
  generalizes to anisotropic media the well-known Cattaneo's relationship 
 \begin{equation}   \label{eq:CATTrel}
	\tau(\theta) \dot{\bf q} + {\bf q} = - \kappa(\theta) grad \theta \, .
\end{equation}

Later  \"{O}nc\"{u}-Moodie \cite{RefOncuMoodie} extended the above papers to the case of a deformable thermoelastic body; there it is shown that
when the referential heat flux ${\bf Q}$, the deformaton gradient ${\bf F}$ and the temperature $\theta$ obey the relation
\begin{equation}   \label{eq:OMrelQ}
	{\bf T}({\bf F}, \, \theta) \dot {\bf Q} + {\bf Q} = -{\bf K}({\bf F}, \, \theta) Grad \theta \, ,
\end{equation}
 with $Grad \theta= {\bf F}^T grad \theta$, and ${\bf T}({\bf F}, \, \theta)$ and ${\bf K}({\bf F}, \, \theta)$ non-singular, then the second law of thermodynamics requires that the specific internal energy $\varepsilon$ and the first Piola-Kirchhoff stress ${\bf S}$ satisfy the relations

\begin{eqnarray}
\rho_R \varepsilon = \rho_R \hat \varepsilon_o({\bf F}, \, \theta) +
 {\bf Q} \cdot {\bf A}({\bf F}, \, \theta) {\bf Q} \, , \\
 {\bf S} = {\bf S}_o({\bf F}, \, \theta) +
 {\bf Q} \cdot {\bf P}({\bf F}, \, \theta) {\bf Q} \, ,
\end{eqnarray}
where $(i)$ $\,\rho_R\,$ is the referential mass density, 
 $(ii)$ $\,\varepsilon_o({\bf F}, \, \theta)\,$ and $\,{\bf S}_o({\bf F}, \, \theta)\,$ are, respectively, the classical specific internal energy per unit mass and Piola-Kirchhoff stress tensor, and  $(iii)$ $\,{\bf A}({\bf F}, \, \theta)$, $\,{\bf P}({\bf F}, \, \theta)\,$
 are defined by
\begin{eqnarray}  \nonumber
	{\bf A}({\bf F}, \, \theta)=- \frac{\theta^2}{2} \frac{\partial}{\partial \theta} \Big[ \frac{{\bf Z}({\bf F}, \, \theta)}{\theta^2} \Big] \, ,\qquad \qquad \qquad \\ 
	{\bf P}({\bf F}, \, \theta)= \frac{1}{2\theta} \frac{\partial}{\partial {\bf F}} {\bf Z}({\bf F}, \, \theta) \, , \qquad
	  {\bf Z}({\bf F}, \, \theta)={\bf K}({\bf F}, \, \theta)^{-1}{\bf T}({\bf F}, \, \theta) \, ; 
\end{eqnarray}
moreover,  $(iv) \,{\bf Z}\,$ is symmetric and {\bf K} is positive-definite.

The present paper extends \"{O}nc\"{u}-Moodie \cite{RefOncuMoodie}  to an electrically polarizable and finitely deformable heat conducting elastic continuum, interacting with the electric field, within a 'second-sound' theory; hence, the constitutive equations include an evolution equation for the heat flux; the latter and the temperature obey a frame-invariant form of Cattaneo's equation.
 In parallel with \cite{RefOncuMoodie}, we derive the thermodynamic restrictions on the constitutive relations and we show that an extended version of the generalized Cattaneo's relation (\ref{eq:OMrelQ}) holds.
 Lastly, in 
 Section \ref{section:On the Theory where the Heat Flux has Response Function} the usual theory is considered, i.e., the one where the heat flux is an independent variable; 
 it is obtained from the previous second-sound theory by deleting the constitutive function for the time rate of the heat 
flux $\dot {\bf q}({\bf X}; t)$ and treating the heat flux as a dependent variable through a constitutive equation.
The thermodynamic restrictions on the constitutive relations and other theorems are found simply by collapsing the corresponding theorems of the second-sound theory.  
\section{Preliminary Definitions}   \label{section:2}
The present paper is an extension of \"{O}nc\"{u}-Moodie \cite{RefOncuMoodie} to thermo-electroelasticity; hence here we treat topics in parallel with \cite{RefOncuMoodie} using quite similar notations.

We consider a body $B$ whose particles are identified with the positions 
$\,{\bf X} \in {\cal E} \,$ they occupy in a fixed reference configuration 
$\,{\cal B}\,$ of a three-dimensional Euclidean point space $\,{\cal E}$. 
A referential mass density
$\, \rho_R(.) : {\cal B} \rightarrow (0, \infty)\,$ is given, so that 
$\, m({\cal P}):=  \int^{}_{{\cal P}}\rho_R dV\,$ is the mass of the part ${\cal P}$ of ${\cal B}$.
We assume that the material filling $B$ is characterized by a given process class $\, I\!\!P(B)\,$ of $B$
as a set of ordered $10-$tuples of functions on ${\cal B} \times I\!\!R$
\begin{equation}     \label{eq:classpr}
	p= \Big( {\bf x}(.), \,  \theta(.), \,  \varphi(.), \,  \varepsilon(.), \, \eta(.) , \, \mbox{\boldmath$\tau$}(.), \,  {\bf P}(.), 
	\,  {\bf q}(.), \,  {\bf b}(.), \,  r(.) \Big) \, \in \, I\!\!P(B) 
\end{equation}
defined with respect to $\,{\cal B}\,$, satisfying the balance laws of linear momentum, moment of momentum, energy, the entropy inequality,and the field equations of electrostatics,
where
\begin{itemize}
	\item $\quad {\bf x}={\bf x}({\bf X}, \, t)\,$ is the {\it motion}, 
	\item $\quad \theta=\theta({\bf X}, \, t)\in (0, \infty)\,$ is the {\it absolute temperature}, 
	\item $\quad \varphi=\varphi({\bf X}, \, t)\,$ is the {\it electric potential}, 
	\item $\quad \varepsilon=\varepsilon({\bf X}, \, t)\,$ is the specific {\it internal energy} per unit mass, 
	\item $\quad \eta=\eta({\bf X}, \, t)\,$ is the specific {\it entropy} per unit mass, 
	\item $\quad \mbox{\boldmath$\tau$}=\mbox{\boldmath$\tau$}({\bf X}, \, t)\, \; \Big({\bf S}={\bf S}({\bf X}, \, t)\Big)\,$ is the {\it Cauchy} ({\it first Piola-Kirchhoff}) {\it stress tensor}, 
	\item $\quad {\bf P}={\bf P}({\bf X}, \, t)\, \; \Big({\bf I\!P}={\bf I\!P}({\bf X}, \, t) \Big)\,$ is the spatial (referential) {\it polarization vector}, 
	\item $\quad {\bf q}={\bf q}({\bf X}, \, t)\, \; \Big({\bf Q}={\bf Q}({\bf X}, \, t)\Big)\,$ is the spatial (referential) {\it heat flux vector}, 
	\item $\quad {\bf b}={\bf b}({\bf X}, \, t)\,$ is the external specific {\it body force} per unit mass,  
	\item $\quad r=r({\bf X}, \, t)\,$ is the {\it radiating heating} per unit mass. 
\end{itemize}

The referential and spatial heat flux and polarization vectors are related by
\begin{equation}     \label{eq:PJP}
{\bf I\!P}=J{\bf F}^{-1}{\bf P}\,, \qquad  {\bf Q}=J{\bf F}^{-1}{\bf q} \, .
\end{equation}
Any motion ${\bf x}(.,.)$ of $B$ is a {\it regular} function, in the sense that it is  continuously differentiable, of class $C^2$ with respect to $t$, and at each time $t$ it is an invertible function from  
$\,{\cal B}\,$ into $\,{\cal E}$.

We use $Grad$, $grad$, $Div$, $div$, to denote gradient and divergence with respect to ${\bf X}$ and ${\bf x}$, respectively, whereas a superposed dot denotes the material time derivative.

The {\it deformation gradient} ${\bf F}$ at ${\bf X}$ at time $t$ is given by
\begin{equation}
	{\bf F}={\bf F}({\bf X}, \, t)=Grad \, {\bf x}({\bf X}, \, t) \, ,
\end{equation}
and the invertibility of the deformation is sured by the condition $$ J=det{\bf F}>0 \, .$$
 
The {\it velocity} ${\bf v}$ of ${\bf X}$ at time $t$ is given by
\begin{equation}
	{\bf v}={\bf v}({\bf X}, \, t)=\dot{\bf x}({\bf X}, \, t) \, .
\end{equation}

The local law of conservation of mass is expressed by
\begin{equation}  \label{eq:cmassM}
	\rho_R = \rho J \, ,   \qquad \dot \rho + \rho div{\bf v}=0 \,,
\end{equation}
where $\;\rho=\rho({\bf X}, \, t)\;$ is the mass density of ${\bf X}$ at time $t$.

The electric potential $\varphi$, together with the polarization vector field ${\bf P}$, in Gaussian units determines the {\it eulerian electric displacement field} by the equality
\begin{equation}    \label{eq:euldisplvectorM}
{\bf D}={\bf E}^M+ 4\pi{\bf P} \, ,
	\end{equation}
	where $\,{\bf E}^M=- \nabla_{_{\bf x}}  \varphi \,$  is the {\it (Maxwellian) spatial electric vector field}.  
	Hence the {\it referential electric vector field} is 
\begin{equation}    \label{eq:refdisplvectorM}
{\bf \Delta}=J{\bf F}^{-1}{\bf D}=J{\bf F}^{-1}{\bf E}^M+ 4\pi{\bf I\!P} \, .
	\end{equation}
	
	The spatial and referential polarization vectors  per unit volume are respectively defined by
\begin{equation}          \label{eq:eulPolarVectorUnVolM}
	{\bf \mbox{\boldmath$\pi$}}={\bf P}/{\rho}  \, , \qquad {\bf \Pi}={\bf I\!P}/{\rho_R} \, .
\end{equation}

Note that any two corresponding referential and spatial 'energy-flux' vectors, that are related by
\begin{equation}    \label{eq:refdisplvectorEnFlM}
{\bf I\!H}=J{\bf F}^{-1}{\bf h}  \,  ,
	\end{equation}
have spatial and referential divergences that are related by

\begin{equation}    \label{eq:refdisplvectorEnFlDIVVM}
DIV {\bf I\!H} = J div{\bf h} 
	\end{equation}

	\section{Spatial Description}
	\subsection{Local balance laws in spatial form}
		\label{subsection:Local balance laws in spatial form}
Under suitable assumptions of regularity, and using (\ref{eq:cmassM}), the usual integral forms 
of the balance laws of linear momentum, moment of momentum, energy, the field equations of electrostatics, and the entropy inequality
are equivalent to the spatial field equations
\begin{equation}        \label{eq:eqmotM}
	\rho \dot{\bf v} = div \mbox{\boldmath$\tau$} + {\bf P} \cdot \nabla_x{\bf E}^M	+ \rho {\bf b} \, , 
\end{equation}  

\begin{equation}     \label{eq:energyM}
	\rho \dot{\varepsilon} = 
	\mbox{\boldmath$\tau$} \cdot \nabla {\bf v} - div{\bf q} + {\bf E}^M \cdot \rho \dot {\bf \mbox{\boldmath$\pi$}} + \rho r\, , 
\end{equation}

\begin{equation}
{\bf E}^M=- \nabla_{_{\bf x}}\varphi  \, , 
\end{equation}

\begin{equation}    \label{eq:entrineqM}
	\rho \dot{\eta} \geq \rho (r/ \theta) - div({\bf q}/\theta) \, . 
\end{equation}

Incidentally, note that by the continuity equation (\ref{eq:cmassM}) we find

 \begin{equation}    \label{eq:PM}
{\bf E}^M \cdot \rho \dot {\bf \mbox{\boldmath$\pi$}}= {\bf E}^M \cdot (\dot {\bf P} + {\bf P} div {\bf v}) \, . 
\end{equation}

Let $\,\psi=\psi()\,$ be the specific {\it free energy} per unit mass defined by
 \begin{equation}    \label{eq:freeenM}
	\psi= \varepsilon - \theta \eta - {\bf E}^M \cdot {\bf \mbox{\boldmath$\pi$}} \, . 
\end{equation}

Then (\ref{eq:energyM}) and (\ref{eq:entrineqM}) yield the {\it reduced dissipation inequality}
 \begin{equation}    \label{eq:DissIneq}
	\rho ( \dot \psi + \eta \dot{\theta}) - 	\mbox{\boldmath$\tau$}\cdot \nabla{\bf v} + \frac{1}{\theta} {\bf q} \cdot {\bf g} 
+ \rho{\bf \mbox{\boldmath$\pi$}} \cdot \dot {\bf E}^M \, \leq \, 0 \, , 
\end{equation}
where $\,{\bf g}=grad\, \theta({\bf X}, \, t)\,$ is the spatial temperature gradient.

We note that Eqs. (\ref{eq:eqmotM}), (\ref{eq:energyM}), (\ref{eq:entrineqM}) and (\ref{eq:freeenM}) respectively coincide with  Eqs.  (3.23), (3.40), (3.43) and (4.2) of \cite{RefTier1}.

\subsection{Spatial Constitutive Assumptions}
\label{subsection:Spatial Constitutive Assumptions}
Let $\,{\cal D}\,$ be an open, simply connected domain consisting of $\,5-$tuples
$\,( {\bf F}, \, \theta, {\bf E}^M, \, {\bf q},\, {\bf g} )$, and assume that 
if $\,( {\bf F}, \, \theta, {\bf E}^M, \, {\bf q},\, {\bf g} ) \in {\cal D}$, then
$\,( {\bf F}, \, \theta, {\bf E}^M, \, {\bf 0},\, {\bf 0} ) \in {\cal D}$.

\begin{assumption} For every $\, p \in I\!\!P(B)\,$  
the specific free energy $\, \psi({\bf X}, \,t)$,
the specific entropy $\, \eta({\bf X}, \,t)$,  
the Cauchy stress tensor $\, {\bf \mbox{\boldmath$\tau$}}({\bf X}, \,t)$,  
the specific polarization vector $\, {\bf P}({\bf X}, \,t)$, 
and the time rate of the heat flux $\, \dot {\bf q}({\bf X}, \,t)\,$ are given by continuously differentiable functions on  $\,{\cal D}\,$ such that
\begin{equation}   \label{eq:a}
	\psi= \overline{\psi}( {\bf F}, \, \theta, {\bf E}^M, \, {\bf q},\, {\bf g} )  \, ,
\end{equation}
	\begin{equation}   \label{eq:b}
	\eta= \overline{\eta}( {\bf F}, \, \theta, {\bf E}^M, \, {\bf q},\, {\bf g} )   \, ,
	\end{equation} 
	\begin{equation}   \label{eq:c}
\mbox{\boldmath$\tau$}= \mbox{\boldmath$\overline{\tau}$}( {\bf F}, \, \theta, {\bf E}^M, \, {\bf q},\, {\bf g} )  \, ,
	\end{equation}
	\begin{equation}   \label{eq:d}
{\bf P}= \overline{{\bf P}}( {\bf F}, \, \theta, {\bf E}^M, \, {\bf q},\, {\bf g} )  \, ,
	\end{equation}
	\begin{equation}   \label{eq:e}
\dot {\bf q}= {\bf h}( {\bf F}, \, \theta, {\bf E}^M, \, {\bf q},\, {\bf g} ) \, .
\end{equation}
Further, the tensors  $\,\partial_{{\bf q}}{\bf h}(.)\,$ and  $\,\partial_{{\bf g}}{\bf h}(.)\,$ 
are non-singular.
\end{assumption}

Of course, once $\, \rho(.)$, $\,\overline{\eta}(.)$,  
$\, \overline{\psi}(.)\,$  and  $\, \overline{{\bf P}}(.)\,$  are known, then equality (\ref{eq:freeenM}) gives the continuously differentiable function $\, \overline{\varepsilon}(.)\,$ determining 
$\, \varepsilon({\bf X}, \,t) \,$ such that 
\begin{equation}   \label{eq:f}
	\varepsilon= \overline{\varepsilon}( {\bf F}, \, \theta, {\bf E}^M, \, {\bf q},\, {\bf g} ) \, .
	\end{equation}
	
	The assumed properties of the {\it heat flux evolution function} ${\bf h}(.)$ indicate that it is invertible for ${\bf q}$ and also for ${\bf g}$.
 The inverse of ${\bf h}(.)$ with respect to ${\bf q}$ is denoted with
\begin{equation}  \label{eq:hh}
		{\bf q}= {\bf h}^{*}( {\bf F}, \, \theta, {\bf E}^M,\, {\bf g}, \, \dot{\bf q} )  \, .
\end{equation}
Note that 
\begin{equation}  \label{eq:DH}
\partial_{{\bf g}}{\bf h}^{*}(.)
=-[\partial_{{\bf q}}{\bf h}]^{-1}(.)\partial_{{\bf g}}{\bf h}(.) \, ,
\end{equation}

so that the tensor $\, \partial_{{\bf g}}{\bf h}^{*}(.) \,$ is also continuous and non-singular.

Also note that the dependence upon $\,{\bf X}\,$ is not written for convenience, but it is implicit and understood when the body is not materially homogeneous.


\subsection{Coleman-Noll Method and Thermodynamic Restrictions}
Given any motion 
$\, {\bf x}({\bf X}, \, t)$, temperature field $\, \theta({\bf X}, \, t)$, electric potential field $\, \varphi({\bf X}, \, t)$
 and any heat flux field $\, {\bf q}({\bf X}, \, t)$,
the constitutive equations (\ref{eq:a})-(\ref{eq:e}) determine $\,e({\bf X}, \, t)$, $\,\eta({\bf X}, \, t)$, $\,\mbox{\boldmath$\tau$}({\bf X}, \, t)$,
$\,{\bf P}({\bf X}, \, t)$, $\,\dot{\bf q}({\bf X}, \, t)$, 
and the local laws (\ref{eq:eqmotM}) and   (\ref{eq:energyM}) determine $\, {\bf b}({\bf X}, \, t)\,$ 
and $\, r({\bf X}, \, t)$.
Hence for any given motion, temperature field,  electric potential field and heat flux field a corresponding process $p$ is constructed.

The method of Coleman-Noll \cite{RefColNoll} is based on the postulate that every process $p$ so constructed belongs to the process class $\,I\!\!P(B)\,$ of $B$, that is, on the assumption that the constitutive assumptions (\ref{eq:a})-(\ref{eq:e}) are compatible with thermodynamics, in the sense of the following

\bigskip 

{\bf Dissipation Principle} {\it $\;$ 
For any given motion, temperature field, electric potential field and heat flux field the process $p$ constructed from  the constitutive equations (\ref{eq:a})-(\ref{eq:e}) belongs to the process  class $\,I\!\!P(B)\,$ of $B$.
Therefore the constitutive functions (\ref{eq:a})-(\ref{eq:e}) are compatible with the second law of thermodynamics in the sense that they satisfy the dissipation inequality
(\ref{eq:entrineqM}). }

	\medskip
	
It is a matter of routine to extend to thermo-electroelasticity Coleman's remark for thermoelasticity written in \cite{RefColNoll},  on page 1119, lines 8-30 from top; such extension, that includes the electric field, is written here just by paraphrasing Coleman. 
 
\begin{remark}          \label{remark:ColMizRem}
Let ${\bf A}(t)$ be any time-dependent invertible tensor, 
$\alpha(t)$ any time-dependent positive scalar,  
${\bf a}(t)$ any time-dependent vector, 
$\beta(t)$ be any time-dependent scalar, ${\bf b}(t)$ any time-dependent vector, and 
$\,Y\,$ any material point of $\,B \,$ whose spatial position in the reference configuration $\,{\bf B}\,$ is ${\bf Y}$.
We can always construct at least one admissible electro-thermodynamic process in 
$\,{\bf B}\,$ such that  
$${\bf F}({\bf X}, \,t ) , \; \theta({\bf X}, \,t ), \; {\bf g}({\bf X}, \,t ),\; {\bf E}^M({\bf X}, \,t ) $$ 
have, respectively, the values
$\,{\bf A}(t), \, \alpha(t),\, {\bf a}, \, {\bf b}\,$ 
at $\, {\bf X}={\bf Y}$.

An example of such a process is the one determined by the following deformatoin function, temperature distribution and electric potential:
\begin{equation}
	{\bf x}={\bf x}({\bf X}, \, t)={\bf Y}+{\bf A}(t)[{\bf X}-{\bf Y}]\,,
\end{equation}
\begin{equation}
	\theta=\theta({\bf X}, \, t)=\alpha(t)+[{\bf A}^T(t){\bf a}(t)] \cdot [{\bf X}-{\bf Y}]\,,
\end{equation}
\begin{equation}
	\varphi=\varphi({\bf X}, \, t)=\beta(t)+[{\bf A}^T(t){\bf b}(t)] \cdot [{\bf X}-{\bf Y}]\,.
\end{equation}
Thus, at a given time  $t$, we can arbitrarily specify not only 
$\,{\bf F}, \; \theta, \; {\bf g}\, $ and  $\,{\bf E}^M \,$
but also their time derivatives
$\,\dot{\bf F} , \; \dot\theta, \; \dot{\bf g}\, $ and  $\,\dot{\bf E}^M\,$ at a point $\,{\bf Y}\,$ and be sure that there exists at least one electro-thermodynamic process corresponding to this choice. 
\end{remark}
 
The next theorem is proved by using Remark \ref{remark:ColMizRem}.

\begin{theorem}   \label{theorem:dissPrSpatial}
The Dissipation Principle is satisfied if and only if the following conditions hold:

(i) the free energy response function 
$\,\overline{\psi}({\bf F}, \, \theta, {\bf E}^M, \, {\bf q},\, {\bf g} )  \,$
is independent of the temperature gradient ${\bf g}$ and determines the entropy,  the first Piola-Kirchhoff stress, and the polarization vector through the relations
\begin{equation}   \label{eq:constRestreta}
\overline{\eta}({\bf F}, \, \theta, {\bf E}^M, \, {\bf q})=
	-\partial_{\theta} \overline{\psi}({\bf F}, \, \theta, {\bf E}^M, \, {\bf q})  \, , 
\end{equation}

\begin{equation}   \label{eq:constRestrK}
\mbox{\boldmath$\overline{\tau}$}({\bf F}, \, \theta, {\bf E}^M, \, {\bf q})=
\rho \partial_{{\bf F}} \overline{\psi}({\bf F}, \, \theta, {\bf E}^M, \, {\bf q})
\end{equation}

\begin{equation}   \label{eq:constRestrP}
\overline{{\bf \mbox{\boldmath$\pi$}}}({\bf F}, \, \theta, {\bf E}^M, \, {\bf q})
 =-\partial_{{\bf E}^M} \overline{\psi}({\bf F}, \, \theta, {\bf E}^M, \, {\bf q})
\end{equation}

(ii) the reduced dissipation inequality
\begin{equation}    \label{eq:RedDissIneq}
\rho \theta \partial_{{\bf q}} \overline{\psi}({\bf F}, \, \theta, {\bf E}^M, \, {\bf q})
\cdot {\bf h}({\bf F}, \, \theta, {\bf E}^M, \, {\bf q},\, {\bf g} ) + 
{\bf q} \cdot {\bf g} \, \leq \, 0 \, 
\end{equation}
is satisfied.  \end{theorem}
\smallskip

\underline{Proof}.  By the chain rule we have
\begin{equation}   \label{eq:psiChainRule}
	\dot \psi = \partial_{{\bf F}} \overline{\psi} \cdot \dot {\bf F}
	            + \partial_{\theta} \overline{\psi} \cdot \dot \theta
	            + \partial_{{\bf E}^M} \overline{\psi} \cdot \dot {\bf E}
	            + \partial_{{\bf q}} \overline{\psi} \cdot \dot {\bf q}
	            + \partial_{{\bf g}} \overline{\psi} \cdot \dot {\bf g}  \, .
\end{equation}

Thus by substituting this equation together with the constitutive equations (\ref{eq:a})-(\ref{eq:e}) into the dissipation inequality (\ref{eq:DissIneq}) gives

\begin{eqnarray}                               \label{eq:ConsPsiChainRule}
	(\rho \partial_{{\bf F}} \overline{\psi} - \mbox{\boldmath$\overline{\tau}$}) \cdot \dot {\bf F}
	+ 	(\rho \partial_{\theta} \overline{\psi} + \overline{\eta}) \dot \theta
	+ 	(\rho \partial_{{\bf E}^M} \overline{\psi} + \overline{{\bf P}}) \cdot \dot {\bf E}^M 
	 \qquad \qquad \\
	+ 	\rho \partial_{{\bf q}} \overline{\psi} \cdot {\bf h}
	+ 	\rho \partial_{{\bf g}} \overline{\psi} \cdot \dot {\bf g} 
	+ \frac{1}{\theta} {\bf q} \cdot {\bf g} 
  \, \leq \, 0 \, .
\end{eqnarray}

Now we proceed following the Coleman-Mizel \cite{RefColNoll} method: by Remark \ref{remark:ColMizRem} we can state that 
$\,\dot {\bf F}, \, \dot \theta, \, \dot {\bf E}^M\,$ and  $\, \dot {\bf g} \,$
can be assigned arbitrary values independently from the other variables and 
the theorem is proved. $\; \diamondsuit$

\medskip

The next theorem extends Theorem 2 of \cite{RefColNoll}.
\begin{theorem}
The time derivative of the heat flux $\, \dot{\bf q}\,$ vanishes for all termal equilibrium states 
$\,( {\bf F}, \, \theta, {\bf E}^M, \, {\bf 0},\, {\bf 0} ) \in {\cal D}\,$ and the tensor
\begin{equation}
{\bf K}({\bf F}, \, \theta, {\bf E}^M)\,=\, 
\partial_{{\bf q}} {\bf h}({\bf F}, \, \theta, {\bf E}^M, \, {\bf 0}, \, {\bf 0})^{-1}
\partial_{{\bf g}} {\bf h}({\bf F}, \, \theta, {\bf E}^M, \, {\bf 0}, \, {\bf 0})
\end{equation}
is positive-definite.
\end{theorem}
\underline{Proof}.  
In the proof of Theorem 2 on p.94 in \cite{RefColNoll} insert 
the additional variable $\,{\bf E}^M\,$ in each occurrence of 
$\,{\bf H}^*, \,{\bf H}, f, \, \,{\bf K}, \,{\bf K}^*\,$; moreover replace $\,{\bf H}^*, \,{\bf H}\,$ with $\,{\bf h}^*, \,{\bf h}$, respectively;
thus we obtain a proof for the present theorem. $\; \diamondsuit$
\subsection{Use of Invariant response functions}
\label{subsection:Use of Invariant response functions}
In order to satisfy the principle of material objectivity $\varepsilon$ and  $\psi$ must be scalar invariant under rigid rotations of the deformed and polarized body.
The invariance of  $\psi$ in a rigid rotation is assured when  $\psi$ is an arbitrary function of the referential quantities $\,E_{LM},\, \theta, \, W_{L}, \, Q_{L}, \, G_{L}$,
where 
\begin{equation}   \label{eq:ERS}
	E_{LM}=\frac{1}{2}(C_{LM}-\delta_{LM}), \qquad   C_{LM}=x_{k,\,L}x_{k,\,M} \, ,
\end{equation}

\begin{equation}  \label{eq:WE}
 W_L=-\frac{\partial \varphi}{\partial X_L} = -\frac{\partial \varphi}{\partial x^p} \frac{\partial x^p}{\partial X_L}\,, \qquad  {\bf W}={\bf F}^{T}{\bf E}^M \, ,
\end{equation}
\begin{equation}   \label{eq:QqGg}
 {\bf Q}=J{\bf F}^{-1}{\bf q}\,, \qquad  {\bf G}={\bf F}^{T}{\bf g} \, .
\end{equation}

Hence we assume that 
\begin{equation}     \label{eq:psi=hatpsi}
	\psi=\tilde \psi( {\bf E}, \, \theta, {\bf W}, \, {\bf Q},\, {\bf G} )  \, .
\end{equation}

Next we calculate the time derivatives in equation (\ref{eq:psiChainRule}) by using $\tilde\psi$ in place of $\overline{\psi}$. 
We find 
\begin{equation} \label{eq:psitime1}
	 \frac{\partial \overline{\psi}}{\partial \bf F}\cdot \dot{\bf F} 
	 = \Big[ \frac{\partial \tilde \psi}{\partial E_{RS}} \, \frac{\partial E_{RS}}{\partial (\partial x^i/\partial X_K)}  
	 + \frac{\partial \tilde \psi}{\partial W_R}  \frac{\partial W_R}{\partial (\partial x^i/\partial X_K)} 
	 + \frac{\partial \tilde \psi}{\partial G_R}  \frac{\partial G_R}{\partial (\partial x^i/\partial X_K)} 
	 \Big] \frac{d}{dt}\frac{\partial x^i}{\partial X_K} \\
\end{equation}

Now, by (\ref{eq:ERS})-(\ref{eq:QqGg}),

\begin{eqnarray} \label{eqnarray:psitime11}  \nonumber
   \frac{\partial \tilde \psi}{\partial E_{RS}} \, \frac{\partial E_{RS}}{\partial (\partial x^i/\partial X_K)}   \frac{d}{dt}\frac{\partial x^i}{\partial X_K}
=
\frac{\partial \tilde \psi}{\partial E_{RS}}
 \frac{1}{2}\Big( \delta_{RK}\frac{\partial x^i}{\partial X_S} + \frac{\partial x^i}{\partial X_R}\delta_{SK}
  \Big) \frac{\partial \dot x^i}{\partial X_K} \\
= \frac{1}{2} \Big( 
   \frac{\partial \tilde \psi}{\partial E_{KS}}\frac{\partial x^i}{\partial X_S} 
   + \frac{\partial \tilde \psi}{\partial E_{RK}}\frac{\partial x^i}{\partial X_R} 
  \Big) \frac{\partial \dot x^i}{\partial X_K}
  =  \frac{\partial \tilde \psi}{\partial E_{RK}}\frac{\partial x^i}{\partial X_R} 
  \frac{\partial \dot x^i}{\partial X_K} = \Big( \frac{\partial \tilde \psi}{\partial {\bf E}}{\bf F}^T \Big) \cdot \dot {\bf F}  \, ,
\end{eqnarray}

\begin{equation}
	\frac{\partial \tilde \psi}{\partial W_R}  \frac{\partial W_R}{\partial (\partial x^i/\partial X_K)}\frac{d}{dt}\frac{\partial x^i}{\partial X_K} =
	\frac{\partial \tilde \psi}{\partial W_R}  \delta_{KR} E^M_i\frac{d}{dt}\frac{\partial x^i}{\partial X_K}=
	\Big( \frac{\partial \tilde \psi}{\partial {\bf W}} \otimes
 {\bf E}^M \Big)\cdot \dot {\bf F} \, ,
\end{equation}
and, similarly with the latter,
\begin{equation}
	\frac{\partial \tilde \psi}{\partial G_R}  \frac{\partial G_R}{\partial (\partial x^i/\partial X_K)}\frac{d}{dt}\frac{\partial x^i}{\partial X_K} =
	\frac{\partial \tilde \psi}{\partial G_R}  \delta_{KR} g_i\frac{d}{dt}\frac{\partial x^i}{\partial X_K}=
	\Big( \frac{\partial \tilde \psi}{\partial {\bf G}} \otimes
 {\bf g} \Big)\cdot \dot {\bf F} \, ;
\end{equation}
	hence 
	
\begin{equation} \label{eq:psitime1111}
	 \frac{\partial \overline{\psi}}{\partial \bf F} \cdot \dot {\bf F}
	 = \Big( \frac{\partial \tilde \psi}{\partial {\bf E}}{\bf F}^T 
	 +  \frac{\partial \tilde \psi}{\partial {\bf W}} \otimes
 {\bf E}^M 
	 + \frac{\partial \tilde \psi}{\partial {\bf G}} \otimes
 {\bf g} \Big)\cdot \dot {\bf F} \, ;
\end{equation}

moreover,
\begin{equation} \label{eq:psitime223}
	 \frac{\partial \overline{\psi}}{\partial \bf g}\cdot \dot{\bf g} 
=
 \frac{\partial \tilde \psi}{\partial G_R}  \frac{\partial G_R}{\partial g^i}\frac{dg^i}{dt}
 =   \frac{\partial \tilde \psi}{\partial G_R}  F^i_R\frac{dg^i}{dt} 
=\Big({\bf F}\frac{\partial \tilde \psi}{\partial \bf G}\Big)\cdot \dot{\bf g} \, ,
\end{equation}
	
\begin{equation} \label{eq:psitime1122}  \nonumber
\frac{\partial \overline{\psi}}{\partial {\bf E}^M}\cdot \dot {\bf E}^M=
 \Big({\bf F} \frac{\partial \tilde \psi}{\partial {\bf W}}\Big) \cdot \dot {\bf E}^M  \, ,
\end{equation}

\begin{equation} \label{eq:psitime2234}
\frac{\partial \overline{\psi}}{\partial \bf q}\cdot \dot{\bf q} 
=
 \Big(\frac{\partial \tilde \psi}{\partial {\bf Q}}  \frac{\partial {\bf Q}}{\partial {\bf {\bf q}}}\Big) \cdot \dot{\bf q} 
 =   \Big( J{\bf F}^{-T} \frac{\partial \tilde \psi}{\partial {\bf Q}}  \Big) \cdot   \dot{\bf q} \,.
\end{equation}

By recollecting the equalities above we can rewrite the {\it dissipation inequality} (\ref{eq:DissIneq}) as
 \begin{eqnarray}    \label{eqnarray:DissIneq2} \nonumber
	\rho \Big[ 
	\Big(\frac{\partial \tilde \psi}{\partial {\bf E}}{\bf F}^T 
	 +  \frac{\partial \tilde \psi}{\partial {\bf W}} \otimes
 {\bf E}^M 
	 + \frac{\partial \tilde \psi}{\partial {\bf G}} \otimes
 {\bf g} \Big)\cdot \dot {\bf F}  \\  \nonumber
	            + \frac{\partial \tilde \psi}{\partial \theta} \cdot \dot \theta
	            + \Big({\bf F} \frac{\partial \tilde \psi}{\partial {\bf W}}\Big) \cdot \dot {\bf E}^M  
	            +  \Big( J{\bf F}^{-T} \frac{\partial \tilde \psi}{\partial {\bf Q}} \Big) \cdot   \dot{\bf q}
	            + \Big({\bf F}\frac{\partial \tilde \psi}{\partial \bf G}\Big)\cdot \dot{\bf g} + \eta \dot{\theta}\Big] \\ 
	            - 	\mbox{\boldmath$\tau$} \cdot \nabla{\bf v} + \frac{1}{\theta} {\bf q} \cdot {\bf g} 
+ \rho{\bf \mbox{\boldmath$\pi$}} \cdot \dot {\bf E}^M \, \leq \, 0 \, .   
\end{eqnarray}

Now we apply the method of Colemann-Noll.

By the arbitrariness of $\,{\bf g}\,$ we have $\,\partial \tilde \psi/{\partial \bf G}=\bf O\,$.

Note that
\begin{equation}  \label{eq:gradvFF}
	   \frac{\partial v^i}{\partial x^j}=  \frac{\partial v^i}{\partial X^K} \frac{\partial X^K}{\partial x^j}
	   =\frac{\partial X^K}{\partial x^j}\frac{d}{dt}\frac{\partial x^i}{\partial X^K} \, , \end{equation}  
that is,
\begin{equation}  \label{eq:gradvFFAbs}
	   \nabla \bf v = {\bf F}^{-T} \dot {\bf F} \, , \end{equation}  
	   and thus
\begin{equation}  \label{eq:gradvFFAbs++}
\mbox{\boldmath$\tau$} \cdot  \nabla \bf v = {\bf F}^{-1} \mbox{\boldmath$\tau$} \cdot \dot{\bf F} \, ; \end{equation}  

	   hence (\ref{eqnarray:DissIneq2}) reduces to
	   
	   \begin{eqnarray}    \label{eqnarray:DissIneq23} \nonumber
	\rho \Big[ 
	\Big( \frac{\partial \tilde \psi}{\partial {\bf E}} {\bf F}^T
	 +  \frac{\partial \tilde \psi}{\partial {\bf W}} \otimes
 {\bf E}^M - \rho^{-1}{\bf F}^{-1} \mbox{\boldmath$\tau$} \Big)\cdot \dot {\bf F}  
 + \Big( \frac{\partial \tilde \psi}{\partial \theta}+\eta \Big) \cdot \dot \theta \\  
	            + \Big({\bf F} \frac{\partial \tilde \psi}{\partial {\bf W}}
	            + {\bf \mbox{\boldmath$\pi$}} \Big) \cdot \dot {\bf E}^M  
	            +  \Big( J{\bf F}^{-T} \frac{\partial \tilde \psi}{\partial {\bf Q}}\Big) \cdot   \dot{\bf q} \Big] 
	           + \frac{1}{\theta} {\bf q} \cdot {\bf g} \, \leq \, 0 \, .   
\end{eqnarray}

By the arbitrariness of the time derivatives $\, \dot {\bf F}$,$\, \dot \theta$, 
$\, \dot {\bf E}^M\,$ and by substituting the constitutive relation (\ref{eq:e}) we find
\begin{equation}   \label{eq:TPsi}
	\rho^{-1}{\bf F}^{-1} \mbox{\boldmath$\tau$} = \frac{\partial \tilde \psi}{\partial {\bf E}} {\bf F}^T
	 +  \frac{\partial \tilde \psi}{\partial {\bf W}} \otimes {\bf E}^M 
\end{equation}
\begin{equation}   \label{eq:etaPsi}
	\eta = - \frac{\partial \tilde \psi}{\partial \theta} \, ,
\end{equation}
\begin{equation}   \label{eq:piPsi}
{\bf \mbox{\boldmath$\pi$}}=-	{\bf F} \frac{\partial \tilde \psi}{\partial {\bf W}} \, ,
\end{equation}
	   \begin{equation}    \label{eq:DissIneq23} \nonumber
	\rho \Big( J{\bf F}^{-T} \frac{\partial \tilde \psi}{\partial {\bf Q}}({\bf E}, \, \theta, {\bf W}, \, {\bf Q})\Big) \cdot  {\bf h}({\bf F}, \, \theta, {\bf E}^M, \, {\bf q}) + \frac{1}{\theta} {\bf q} \cdot {\bf g} \, \leq \, 0 \, .   
\end{equation}

Of course, by (\ref{eq:psitime2234}) the latter inequality rewrites as
	   \begin{equation}    \label{eq:DissIneq23REW} \nonumber
\rho \Big( \frac{\partial \overline{\psi}}{\partial {\bf q}}({\bf F}, \, \theta, {\bf E}^M, \, {\bf q})\Big) \cdot  {\bf h}({\bf F}, \, \theta, {\bf E}^M, \, {\bf q}) + \frac{1}{\theta} {\bf q} \cdot {\bf g} \, \leq \, 0 \, .   
\end{equation}

We have proved the version of Theorem \ref{theorem:dissPrSpatial} that employes the objective energy response function $\, \tilde \psi$.

\begin{theorem}   \label{theorem:dissPrSpatialHAT}
The Dissipation Principle is satisfied if and only if the following conditions hold:

(i) the objective free energy response function 
$\,\tilde \psi( {\bf E}, \, \theta, {\bf W}, \, {\bf Q},\, {\bf G} )  \,$
is independent of the temperature gradient ${\bf G}$ and determines the entropy,  the Cauchy stress tensor, and the polarization vector per unit mass through the relations (\ref{eq:TPsi})-(\ref{eq:piPsi});

(ii) the reduced dissipation inequality (\ref{eq:DissIneq23})
is satisfied.  
  \end{theorem}

We point out that Equalities (\ref{eq:TPsi}), (\ref{eq:eulPolarVectorUnVolM}) and (\ref{eq:piPsi}) yield the following expression for the Cauchy stress: 
\begin{equation}   \label{eq:TPsiExpr}
	\mbox{\boldmath$\tau$} = \rho {\bf F} \frac{\partial \tilde \psi}{\partial {\bf E}} {\bf F}^T
	 -  {\bf P} \otimes {\bf E}^M \, ;
\end{equation}
hence for the antisymmetric portion $\,\mbox{\boldmath$\tau$}^A\,$  of $\,\mbox{\boldmath$\tau$}\,$ we obtain the expression
\begin{equation}   \label{eq:TPsiExprSKW}
\mbox{\boldmath$\tau$}^A = \frac{1}{2} \,\Big( {\bf E}^M \otimes {\bf P} -  {\bf P} \otimes {\bf E}^M \Big) \, ,
\end{equation}
that coincides with (3.24) of \cite{RefTier1}.


\section{Internal Dissipation and Entropy Equality}   
\label{section:Internal Dissipation and Entropy Equality}  
The local {\it internal dissipation} $\,\delta_o\,$ in a thermoelastic body is defined by (\cite{RefTruesdell}, p.112)
\begin{equation}\label{eq:IntDissip}
\delta_o=\theta \dot \eta -(r -\frac{1}{\rho}div{\bf q}) \, ;
\end{equation}
then one proves that $\, \delta_o \equiv 0 \,$ alond every local thermoelastic process. Within thermo-electroelasticity here we define the {\it internal dissipation} just by (\ref{eq:IntDissip}) and hence we extend the afore-mentioned theorem by the following
\begin{theorem}\label{theorem:diss}
Along any local process of $B$ we have
\begin{equation}\label{eq:Thdiss}
	\delta_o= - \rho \frac{\partial \psi}{\partial {\bf q}} \cdot \dot {\bf q} \geq 
\frac{1}{\theta}\,{\bf q} \cdot {\bf g}  \, .
\end{equation}
\end{theorem}
\underline{Proof}.  
By inserting the energy equation (\ref{eq:energyM}) in (\ref{eq:IntDissip}) we obtain the equality 
\begin{equation}\label{eq:disssec}
	\delta_o=\theta \dot \eta 
-\frac{1}{\rho}\big(\rho \dot \varepsilon - \mbox{\boldmath$\tau$}\cdot \nabla {\bf v} 
          - {\bf E}^M \cdot {\bf \mbox{\boldmath$\pi$}} \big)   \end{equation}

          Now by (\ref{eq:freeenM}) we find 
          
\begin{equation}
	\theta \dot \eta 
	= - \dot \psi + \dot \varepsilon - \dot \theta \eta 
	- \dot {\bf E}^M \cdot {\bf \mbox{\boldmath$\pi$}} 
	- {\bf E}^M \cdot \dot {\bf \mbox{\boldmath$\pi$}} 
\end{equation}
and by replacing the latter into (\ref{eq:disssec}) by the constitutive restrictions 
(\ref{eq:constRestreta})-(\ref{eq:RedDissIneq})  we find (\ref{eq:Thdiss}).


In thermoelasticity one shows that any thermoelastic process is locally reversible, in the sense that the following entropy equality 
\begin{equation}
	\rho \dot \eta = \rho \frac{r}{\theta} - \frac{div{\bf q}}{\theta} \,  
\end{equation}
holds.
 Here this result is generalized to thermo-electroelasticity  by the following
\begin{theorem}\label{theorem:LocRev}
Along any local process of $B$ the following entropy equality
\begin{equation}\label{eq:EntEq}
	\rho \dot \eta =  \rho \frac{r}{\theta} - \frac{div{\bf q}}{\theta} 
	        -  \rho \frac{\partial \psi}{\partial {\bf q}} \cdot \dot {\bf q}\, .
\end{equation}
holds.
\end{theorem}

In words the theorem can be interpreted by the assertion that in thermo-electroelasticity every process is locally reversible. 

\underline{Proof}. Equality (\ref{eq:IntDissip}) is equivalent to (\ref{eq:EntEq}). $\; \diamondsuit$

\section{Referential description}
\label{section:Referential description}
We can rewrite the constitutive relations (\ref{eq:a})-(\ref{eq:e}) in material form by using the first Piola-Kirchhoff stress tensor $\, {\bf S}({\bf X}, \,t)$, that is related with the Cauchy stress by

\begin{equation}
	{\bf S}=J{\bf F}^{-1}\mbox{\boldmath$\tau$} \, 
\end{equation}
and by using the well known equalities (\ref{eq:PJP}), (\ref{eq:refdisplvectorM})-(\ref{eq:refdisplvectorEnFlDIVVM}) and
\begin{equation}
	Div{\bf S}=Jdiv\mbox{\boldmath$\tau$} \, ,   
\end{equation}
Now the process class $\, I\!\!P(B)\,$ of $B$ of Section \ref{section:2}, contatining the processes (\ref{eq:classpr}), must be substituted with  $\,I\!\!P_R(B)$, that is the 
set of ordered $10-$tuples of functions on ${\cal B} \times I\!\!R$
\begin{equation}
	p_R= \Big( {\bf x}(.), \,  \theta(.), \,  \varphi(.), \,  \varepsilon(.), \, \eta(.) , \,  {\bf S}(.), \,  {\bf I\!P}(.), 
	\,  {\bf Q}(.), \,  {\bf b}(.), \,  r(.) \Big) \, \in \, I\!\!P_R(B) 
\end{equation}
defined with respect to $\,{\cal B}\,$, satisfying the material versions of the balance laws of linear momentum, moment of momentum, energy, the entropy inequality,and the field equations of electrostatics.

	\subsection{Local balance laws in material form}
	\label{subsection:Local balance laws in material form}
The local field laws (\ref{eq:eqmotM})-(\ref{eq:entrineqM}) in the referential description write as 
 \begin{equation}        \label{eq:eqmot}
	\rho_R \dot{\bf v} = Div {\bf S} + 	\rho_R {\bf b} \, , 
\end{equation}  

 \begin{equation}     \label{eq:energy}
	\rho_R \dot{\varepsilon} = 
	{\bf S} \cdot \dot{\bf F} - Div{\bf Q} + {\bf W} \cdot \dot {\bf I\!P} + \rho_R r\, , 
\end{equation}

  \begin{equation}
{\bf W}=- \nabla_{_{\bf X}}\varphi \;(=-{\bf F}^T \nabla_{_{\bf x}}\varphi )  \,  , \qquad Div {\bf \Delta} = 0 \, , 
\end{equation}

 \begin{equation}    \label{eq:entrineq}
	\rho_R \dot{\eta} \geq \rho_R (r/ \theta) - Div({\bf Q}/\theta) \, . 
\end{equation}

Let $\,\psi=\psi(.)\,$ be the specific {\it free energy} per unit mass defined by
 \begin{equation}    \label{eq:freeen}
	\psi= \varepsilon - \theta \eta -  {\bf W} \cdot {\bf \Pi} \, . 
\end{equation}

Then (\ref{eq:energy}) and (\ref{eq:entrineq}) yield the {\it reduced dissipation inequality}
 \begin{equation}    \label{eq:DissIneqM}
	\rho_R ( \dot \psi + \eta \dot{\theta}) - 	{\bf S} \cdot \dot{\bf F} +  \frac{1}{\theta} {\bf Q} \cdot {\bf G} 
+ {\bf I\!P} \cdot \dot {\bf W} \, \leq \, 0 \, , 
\end{equation}
where $\,{\bf G}=Grad\, \theta({\bf X}, \, t)\,$ is the referential temperature gradient.

\begin{remark}
Note that the free-energy function $\psi$ defined here by (\ref{eq:freeen}) coincides with the analogous function $\chi$ defined in (4.2) of 
\cite{RefTier1}, on page 596.
In fact, 
$${\bf W} \cdot {\bf \Pi}= {\bf W} \cdot {\bf I\!P}/\rho_R 
= ({\bf F}^T {\bf E}^M) \cdot (J{\bf F}^{-1}{\bf P})/\rho_R
={\bf E}^M \cdot {\bf P} (J/\rho_R) \, $$ 
Hence, by $\,\rho_R=J\rho$, we have
$${\bf W} \cdot {\bf \Pi}
={\bf E}^M \cdot {\bf P}/\rho
= {\bf E}^M \cdot {\bf \mbox{\boldmath$\pi$}}\, ,$$
where $\, \mbox{\boldmath$\pi$}\,$ is the spatial polarization vector. 
\end{remark}

\subsection{Referential Constitutive Assumptions}
\label{subsection:Referential Constitutive Assumptions}
Let $\,{\cal D}_R\,$ be the open, simply connected domain consisting of $\,5-$tuples
$\;( {\bf F}, \, \theta, {\bf W}, \, {\bf Q},\, {\bf G} )\,$  such that
$\,( {\bf F}, \, \theta, {\bf W}, \, {\bf Q},\, {\bf G} ) \in {\cal D}$;
hence if $\,( {\bf F}, \, \theta, {\bf W}, \, {\bf Q},\, {\bf G} ) \in {\cal D}_R$, then
$\,( {\bf F}, \, \theta, {\bf W}, \, {\bf 0},\, {\bf 0} )  \in  {\cal D}_R\,$.

In order to extend the topics of ONCU, next we use a free energy function of the form
\begin{equation}     \label{eq:psi=overpsi}
	\psi=\hat{\psi}( {\bf F}, \, \theta, {\bf W}, \, {\bf Q},\, {\bf G} )  \, .
\end{equation}
\begin{assumption} For every $\, p \in I\!\!P_R(B)\,$  
the specific free energy $\, \psi({\bf X}, \,t)$,
the specific entropy $\, \eta({\bf X}, \,t)$,  
the first Piola-Kirchhoff stress tensor $\, {\bf S}({\bf X}, \,t)$,  
the specific polarization vector $\, {\bf I\!P}({\bf X}, \,t)$, 
and the time rate of the heat flux $\, \dot {\bf Q}({\bf X}, \,t)\,$ are given by continuously differentiable functions on  $\,{\cal D}_R\,$ such that
\begin{equation}   \label{eq:aM}
	\psi= \hat{\psi}( {\bf F}, \, \theta, {\bf W}, \, {\bf Q},\, {\bf G} )  \, ,
\end{equation}
	\begin{equation}   \label{eq:bM}
	\eta= \hat{\eta}( {\bf F}, \, \theta, {\bf W}, \, {\bf Q},\, {\bf G} )   \, ,
	\end{equation} 
	\begin{equation}   \label{eq:cM}
	{\bf S}= \hat{{\bf S}}( {\bf F}, \, \theta, {\bf W}, \, {\bf Q},\, {\bf G} )  \, ,
	\end{equation}
	\begin{equation}   \label{eq:dM}
{\bf I\!P}= \hat{{\bf I\!P}}( {\bf F}, \, \theta, {\bf W}, \, {\bf Q},\, {\bf G} )  \, ,
	\end{equation}
	\begin{equation}   \label{eq:eM}
\dot {\bf Q}= {\bf H}({\bf F}, \, \theta, {\bf W}, \, {\bf Q},\, {\bf G} ) \, .
\end{equation}
Further, the tensors  $\,\partial_{{\bf Q}}{\bf H}(.)\,$ and  $\,\partial_{{\bf G}}{\bf H}(.)\,$ 
are non-singular.
\end{assumption}

Of course, once $\, \rho_R(.)$, $\, \hat {\bf I\!P}(.)$,  $\, \hat \psi(.)\,$  and  $\, \hat \eta(.)\,$  are known, then equality (\ref{eq:freeen}) gives the continuously differentiable function $\, \hat \varepsilon(.)\,$ determining 
$\, \varepsilon({\bf X}, \,t) \,$ such that 
\begin{equation}   \label{eq:fM}
	\varepsilon= \hat \varepsilon( {\bf F}, \, \theta, {\bf W}, \, {\bf Q},\, {\bf G} )  \, .
	\end{equation}
	
	The assumed properties of the {\it heat flux evolution function} ${\bf H}(.)$ indicate that it is invertible for ${\bf Q}$ and also for ${\bf G}$.
	We denote the inverse of ${\bf H}(.)$ with respect to ${\bf Q}$ with
	
\begin{equation}  \label{eq:H}
		{\bf Q}= {\bf H}^{*}( {\bf F}, \, \theta, {\bf W},\, {\bf G}, \, \dot{\bf Q} )  \, .
\end{equation}

Note that 
\begin{equation}  \label{eq:DHM}
\partial_{{\bf G}}{\bf H}^{*}(.)
=-[\partial_{{\bf Q}}{\bf H}]^{-1}(.)\partial_{{\bf G}}{\bf H}(.) \, ,
\end{equation}
so that the tensor $\, \partial_{{\bf G}}{\bf H}^{*}(.) \,$ is also continuous and non-singular.
Also note that the dependence upon $\,{\bf X}\,$ is not written for convenience, but it is implicit and understood when the body is not materially homogeneous.


\subsection{Coleman-Noll Method and Thermodynamic Restrictions}
\label{subsection:Coleman-Noll Method and Thermodynamic Restrictions}
Given any motion 
$\, {\bf x}({\bf X}, \, t)$, temperature field $\, \theta({\bf X}, \, t)$, electric potential field $\, \varphi({\bf X}, \, t)$
 and any heat flux field $\, {\bf Q}({\bf X}, \, t)$,
the constitutive equations (\ref{eq:aM})-(\ref{eq:eM}) determine $\,e({\bf X}, \, t)$, $\,\eta({\bf X}, \, t)$, $\,{\bf S}({\bf X}, \, t)$,
$\,{\bf I\!P}({\bf X}, \, t)$, $\,\dot{\bf Q}({\bf X}, \, t)$, 
and the local laws (\ref{eq:eqmot}) and   (\ref{eq:energy}) determine $\, {\bf b}({\bf X}, \, t)\,$ 
and $\, r({\bf X}, \, t)$.
Hence for any given motion, temperature field,  electric potential field and  heat flux field a corresponding process $p$ is constructed.

The method of Coleman-Noll \cite{RefColNoll} is based on the postulate that every process $p$ so constructed belongs to the process class $\,I\!\!P(B)\,$ of $B$, that is, on the assumption that the constitutive assumptions (\ref{eq:aM})-(\ref{eq:eM}) are compatible with thermodynamics, in the sense of the following

\bigskip 

{\bf Dissipation Principle} {\it $\;$ 
For any given motion, temperature field, electric potential field and related heat flux field the process $p$ constructed from  the constitutive equations (\ref{eq:aM})-(\ref{eq:eM}) belongs to the process  class $\,I\!\!P(B)\,$ of $B$.
Therefore the constitutive functions (\ref{eq:aM})-(\ref{eq:eM}) are compatible with the second law of thermodynamics in the sense that they satisfy the dissipation inequality
(\ref{eq:entrineq}). }

\begin{theorem}
The Dissipation Principle is satisfied if and only if the following conditions hold:

(i) the free energy response function 
$\,\hat \psi( {\bf F}, \, \theta, {\bf W}, \, {\bf Q},\, {\bf G} )  \,$
is independent of the temperature gradient ${\bf G}$ and determines the entropy,  the first Piola-Kirchhoff stress, and the polarization vector through the relations
\begin{equation}   \label{eq:constRestretaM}
\hat\eta({\bf F}, \, \theta, {\bf W}, \, {\bf Q})=
	-\partial_{\theta} \hat \psi({\bf F}, \, \theta, {\bf W}, \, {\bf Q})  \, , 
\end{equation}

\begin{equation}   \label{eq:constRestrKM}
\hat{\bf S}({\bf F}, \, \theta, {\bf W}, \, {\bf Q})=
\rho_R \partial_{{\bf F}} \hat \psi({\bf F}, \, \theta, {\bf W}, \, {\bf Q})
\end{equation}

\begin{equation}   \label{eq:constRestrPM}
\hat{\bf \Pi}({\bf F}, \, \theta, {\bf W}, \, {\bf Q})
 =-\partial_{{\bf W}} \hat \psi({\bf F}, \, \theta, {\bf W}, \, {\bf Q})
\end{equation}

(ii) the reduced dissipation inequality
\begin{equation}    \label{eq:RedDissIneqM}
\rho_R \theta \partial_{{\bf Q}} \hat \psi({\bf F}, \, \theta, {\bf W}, \, {\bf Q})
\cdot \hat {\bf H}({\bf F}, \, \theta, {\bf W}, \, {\bf Q},\, {\bf G} ) + 
{\bf Q} \cdot {\bf G} \, \leq \, 0 \, 
\end{equation}
is satisfied.  \end{theorem}
\smallskip

\underline{Proof}.  By he chain rule we have
\begin{equation}   \label{eq:psiChainRuleM}
	\dot \psi = \partial_{{\bf F}} \hat \psi \cdot \dot {\bf F}
	            + \partial_{\theta} \hat \psi \cdot \dot \theta
	            + \partial_{{\bf W}} \hat \psi \cdot \dot {\bf W}
	            + \partial_{{\bf Q}} \hat \psi \cdot \dot {\bf Q}
	            + \partial_{{\bf G}} \hat \psi \cdot \dot {\bf G}  \, .
\end{equation}

Thus by substituting this equation together with the constitutive equations (\ref{eq:aM})-(\ref{eq:eM}) into the dissipation inequality (\ref{eq:DissIneqM}) gives

\begin{eqnarray}                               \label{eq:ConsPsiChainRuleM}
	(\rho_R \partial_{{\bf F}} \hat \psi - \hat {\bf S}) \cdot \dot {\bf F}
	+ 	(\rho_R \partial_{\theta} \hat \psi + \hat \eta) \dot \theta
	+ 	(\rho_R \partial_{{\bf W}} \hat \psi + \hat {\bf I\!P}) \cdot \dot {\bf W} 
	 \qquad \qquad \\
	+ 	\rho_R \partial_{{\bf Q}} \hat \psi \cdot {\bf H}
	+ 	\rho_R \partial_{{\bf G}} \hat \psi \cdot \dot {\bf G} 
	+ \frac{1}{\theta} {\bf Q} \cdot {\bf G} 
  \, \leq \, 0 \, .
\end{eqnarray}
Now we follow the Coleman-Mizel \cite{RefColNoll} method: by Remark \ref{remark:ColMizRem}, written in referential form, we can state that 
$\,\dot {\bf F}, \, \dot \theta, \, \dot {\bf W}\,$ and  $\, \dot {\bf G} \,$
can be assigned arbitrary values independently from the other variables; thus
the theorem is proved. $\, \diamondsuit$

 Simply by inserting the additional variable $\,{\bf W}\,$
in each occurrence of 
$\,{\bf H}^*, \,{\bf H}, f, \, \,{\bf K}, \,{\bf K}^*\,$ in
Theorem $2$ of \cite{RefColNoll} and in its proof,
we obtain that the proof of the theorem below.
\begin{theorem}
The time derivative of the heat flux $\, \dot{\bf Q}\,$ vanishes for all termal equilibrium states 
$\,( {\bf F}, \, \theta, {\bf W}, \, {\bf 0},\, {\bf 0} ) \in {\cal D}\,$ and the tensor
\begin{equation}
{\bf K}({\bf F}, \, \theta, {\bf W})\,=\, 
\partial_{{\bf Q}} {\bf H}({\bf F}, \, \theta, {\bf W}, \, {\bf 0}, \, {\bf 0})^{-1}
\partial_{{\bf G}} {\bf H}({\bf F}, \, \theta, {\bf W}, \, {\bf 0}, \, {\bf 0})
\end{equation}
is positive-definite.
\end{theorem}

\section{On Cattaneo's Equation}

The results and considerations in NN.4, 5 of \cite{RefColNoll} remain true also in the context of the present paper.  Of course one has to add the electric field $\,{\bf W}\,$ as an independent variable in the arguments of each constitutive quantity.  Since this vector is referential, it does not change under a change of observer.
Hence the following theorem holds, where it is assumed that 
$\,\hat {\bf H}({\bf F}, \, \theta, {\bf W}, \, {\bf Q},\, {\bf G} ) \,$ is linear in $\,{\bf Q}\,$ and $\,{\bf G}$.

\begin{theorem}  \label{theorem:catt}
Let the evolution equation of the heat flux be given by the following form of Cattaneo's equation:
\begin{equation}   \label{eq:CFOrelWW}
	\hat{\bf T}({\bf F}, \, \theta,\, {\bf W}) \dot {\bf Q} + {\bf Q} = - \hat{\bf K}({\bf F}, \, \theta,\, {\bf W}) {\bf G} \, .
\end{equation}
Then the Dissipation Principle is equivalent to the conditions:

(i) the tensor $\, \hat{\bf K}({\bf F}, \, \theta,\, {\bf W}) \, $
is positive definite;

(ii)  the tensor $\, \hat{\bf Z}({\bf F}, \, \theta,\, {\bf W}) \, $
is symmetric;

(iii) the response functions of the specific free energy, specific internal energy, specific entropy and first Piola-Kirchhoff stress are given by
\begin{equation}   \label{eq:aMWW}
	\rho_R \hat{\psi}( {\bf F}, \, \theta, {\bf W},\, {\bf Q} )
	=\rho_R \hat{\psi}_o( {\bf F}, \, \theta, {\bf W}) + 
	\frac{1}{2 \theta}{\bf Q} \cdot \hat{\bf Z}({\bf F}, \, \theta,\, {\bf W}){\bf Q} \, ,
\end{equation}
\begin{equation}   \label{eq:bMWW}
	\rho_R \hat{\varepsilon}( {\bf F}, \, \theta, {\bf W},\, {\bf Q} )
	=\rho_R \hat{\varepsilon}_o( {\bf F}, \, \theta, {\bf W}) + 
	{\bf Q} \cdot \hat{\bf A}({\bf F}, \, \theta,\, {\bf W}){\bf Q} \, ,
\end{equation}
\begin{equation}   \label{eq:cMWW}
	\rho_R \hat{\eta}({\bf F}, \, \theta, {\bf W},\, {\bf Q} )
	=\rho_R \hat{\eta}_o({\bf F}, \, \theta, {\bf W}) + 
	{\bf Q} \cdot \hat{\bf B}({\bf F}, \, \theta,\, {\bf W}){\bf Q} \, ,
\end{equation}
\begin{equation}   \label{eq:dMWW}
	\hat{{\bf S}}( {\bf F}, \, \theta, {\bf W},\, {\bf Q} )
	= \hat{{\bf S}}_o( {\bf F}, \, \theta, {\bf W}) + 
	{\bf Q} \cdot \hat{\bf P}({\bf F}, \, \theta,\, {\bf W}){\bf Q} \, ,
\end{equation}
where 
\begin{equation}   \label{eq:psi0}
	\hat{\psi}_o( {\bf F}, \, \theta, {\bf W}) 
	= \hat{\psi}( {\bf F}, \, \theta, \, {\bf W},\, {\bf 0})  \, ,
\end{equation} 
 \begin{equation}   \label{eq:psi01}
 \hat{\varepsilon}_o( {\bf F}, \, \theta, {\bf W}) 
	= \hat{\psi}_o( {\bf F}, \, \theta, {\bf W}) - \theta \partial_\theta \hat{\psi}_o( {\bf F}, \, \theta, {\bf W})   \, ,
\end{equation}  
 \begin{equation}   \label{eq:psi02}
 \hat{\eta}_o( {\bf F}, \, \theta, {\bf W}) 
	=  - \partial_\theta \hat{\psi}_o( {\bf F}, \, \theta, {\bf W})   \, ,
\end{equation} 
 \begin{equation}   \label{eq:psi03}
 \hat{{\bf S}}_o( {\bf F}, \, \theta, {\bf W}) 
	= \rho_R \partial_{\bf F} \hat{\psi}_o( {\bf F}, \, \theta, {\bf W})   \, ,
\end{equation}
 \begin{equation}   \label{eq:psi04}
 \hat{{\bf Z}}({\bf F}, \, \theta, {\bf W}) 
	= \hat{{\bf K}}( {\bf F}, \, \theta, {\bf W})^{-1} \hat{{\bf T}}( {\bf F}, \, \theta, {\bf W})   \, ,
\end{equation}
 \begin{equation}   \label{eq:psi05}
 \hat{{\bf B}}({\bf F}, \, \theta, {\bf W}) 
	= -\frac{1}{2} \frac{\partial}{\partial \theta}
	 \Big[ \frac{\hat{{\bf Z}}({\bf F}, \, \theta, {\bf W})}{\theta} \Big] \, ,
 \end{equation}
 \begin{equation}   \label{eq:psi06}
 \hat{{\bf P}}({\bf F}, \, \theta, {\bf W})=\frac{1}{2 \theta}
\frac{\partial}{\partial {\bf F}}\hat{\bf Z}({\bf F}, \, \theta, {\bf W})
 \, .\end{equation}
\end{theorem}

Just as it is suggested in  \cite{RefOncuMoodie}, from \cite{RefColFabOwen1} we may call $\,\hat{{\bf K}}( {\bf F}, \, \theta, {\bf W})\,$ the  
{\it steady-state conductivity},
 $\,\hat{{\bf T}}( {\bf F}, \, \theta, {\bf W})\,$ 
 the  
{\it tensor of relaxation times}, and
 $\,\hat{{\bf Z}}^{-1}\,$ 
 the  
{\it instantaneous conductivity}.
\section{On the Theory where the Heat Flux has Response Function}
\label{section:On the Theory where the Heat Flux has Response Function}
Next we consider the (usual) theory where there is no constitutive equation for the {\it rate of heat flux} and the heat flux is treated as a dependent variable. 
In parallel with  \cite{T:NLETE}, \cite{Y:ESF} we follow the method of Coleman-Noll and find the thermodynamic restrictions on the constitutive relations of an electrically polarizable and finitely deformable heat conducting elastic continuum, interacting with the electric field.

We proceed by listing the results in Sections 1 to 6 and by showing where and how they change.  
Sections \ref{section:2} and \ref{subsection:Local balance laws in spatial form} remain unchanged. 
Section \ref{subsection:Spatial Constitutive Assumptions} must be replaced by the section below.
\subsection{Spatial Constitutive Assumptions (usual theory)}
\label{subsection:Constitutive Assumptions in Spatial FormT}
Let $\,{\cal D}\,$ be an open, simply connected domain consisting of $\,4-$tuples
$\,( {\bf F}, \, \theta, {\bf E}^M, \, {\bf g} )$, and assume that 
if $\,( {\bf F}, \, \theta, {\bf E}^M, \, {\bf g} ) \in {\cal D}$, then
$\,( {\bf F}, \, \theta, {\bf E}^M, \, {\bf 0} ) \in {\cal D}$.
\begin{assumption} For every $\, p \in I\!\!P(B)\,$  
the specific free energy $\, \psi({\bf X}, \,t)$,
the specific entropy $\, \eta({\bf X}, \,t)$,  
the Cauchy stress tensor 
$\, \mbox{\boldmath$\tau$}({\bf X}, \, t)$,
the specific polarization vector $\, {\bf P}({\bf X}, \,t)$, 
and the heat flux $\, {\bf q}({\bf X}, \,t)\,$ are given by continuously differentiable functions on  $\,{\cal D}\,$ such that
\begin{equation}   \label{eq:aa}
	\psi= \overline{\psi}( {\bf F}, \, \theta, {\bf E}^M, \, {\bf g} )  \, ,
\end{equation}
	\begin{equation}   \label{eq:bb}
	\eta= \overline{\eta}( {\bf F}, \, \theta, {\bf E}^M, \, {\bf g} )   \, ,
	\end{equation} 
	\begin{equation}   \label{eq:cc}
\mbox{\boldmath$\tau$}= \mbox{\boldmath$\overline{\tau}$}( {\bf F}, \, \theta, {\bf E}^M, \, {\bf g} )  \, ,
	\end{equation}
	\begin{equation}   \label{eq:dd}
{\bf P}= \overline{{\bf P}}( {\bf F}, \, \theta, {\bf E}^M, \, {\bf g} )  \, ,
	\end{equation}
	\begin{equation}   \label{eq:ee}
{\bf q}= \overline{{\bf q}}( {\bf F}, \, \theta, {\bf E}^M, \, {\bf g} ) \, .
\end{equation}
\end{assumption}
Of course, once $\, \rho(.)$, $\, \overline{{\bf P}}(.)$,  $\, \overline{\psi}(.)\,$  and  $\, \overline{\eta}(.)\,$  are known, then equality (\ref{eq:freeen}) gives the continuously differentiable function $\, \overline{\varepsilon}(.)\,$ determining 
$\, \varepsilon({\bf X}, \,t) \,$ such that 
\begin{equation}   \label{eq:ff}
	\varepsilon= \overline{\varepsilon}( {\bf F}, \, \theta, {\bf E}^M, \, {\bf g} ) \, .
	\end{equation}
Also note that the dependence upon $\,{\bf X}\,$ is not written only for brevity; when the body is not materially homogeneous it 	
becomes active.
\subsection{Coleman-Noll Method and Thermodynamic Restrictions (usual theory)}
In Section \ref{subsection:Spatial Constitutive Assumptions} must be replaced the constitutive law (\ref{eq:e}) with  (\ref{eq:ee}).  Thus the sections rewrites as
\begin{assumption} For every $\, p \in I\!\!P(B)\,$  
the specific free energy $\, \psi({\bf X}, \,t)$,
the specific entropy $\, \eta({\bf X}, \,t)$,  
the Cauchy stress tensor $\, \mbox{\boldmath$\tau$}({\bf X}, \, t)$ 
the specific polarization vector $\, {\bf P}({\bf X}, \,t)$, 
and the heat flux $\, {\bf q}({\bf X}, \,t)\,$ are given by continuously differentiable functions on  $\,{\cal D}\,$ such that
\begin{equation}   \label{eq:aT}
	\psi= \overline{\psi}( {\bf F}, \, \theta, {\bf E}^M, \, {\bf g} )  \, ,
\end{equation}
	\begin{equation}   \label{eq:bT}
	\eta= \overline{\eta}( {\bf F}, \, \theta, {\bf E}^M, \, {\bf g} )   \, ,
	\end{equation} 
	\begin{equation}   \label{eq:cT}
\mbox{\boldmath$\tau$}= \mbox{\boldmath$\overline{\tau}$}( {\bf F}, \, \theta, {\bf E}^M, \, {\bf g} )  \, ,
	\end{equation}
	\begin{equation}   \label{eq:dT}
{\bf P}= \overline{{\bf P}}( {\bf F}, \, \theta, {\bf E}^M, \, {\bf g} )  \, ,
	\end{equation}
	\begin{equation}   \label{eq:eT}
{\bf q}= \overline{{\bf q}}( {\bf F}, \, \theta, {\bf E}^M, \, {\bf g} ) \, .
\end{equation}
\end{assumption}

Of course, once $\, \rho(.)$, $\, \overline{{\bf P}}(.)$,  $\, \overline{\psi}(.)\,$  and  $\, \overline{\eta}(.)\,$  are known, then equality (\ref{eq:freeen}) gives the continuously differentiable function $\, \overline{\varepsilon}(.)\,$ determining 
$\, \varepsilon({\bf X}, \,t) \,$ such that 
\begin{equation}   \label{eq:fT}
	\varepsilon= \overline{\varepsilon}( {\bf F}, \, \theta, {\bf E}^M, \, {\bf g} ) \, .
	\end{equation}
	
Again the dependence upon $\,{\bf X}\,$ is not written only for brevity.

\bigskip 
The {\it dissipation principle} and Remark \ref{remark:ColMizRem} must be understood here just as in Section \ref{subsection:Coleman-Noll Method and Thermodynamic Restrictions}; then Theorem \ref{theorem:dissPrSpatial} becomes the one below, whose proof has the same steps in the proof of the 
former by dropping there ${\bf h}$.

\begin{theorem}   \label{theorem:dissPrSpatial00}
The Dissipation Principle is satisfied if and only if the following conditions hold:

(i) the free energy response function 
$\,\overline{\psi}({\bf F}, \, \theta, {\bf E}^M, \,  {\bf g} )  \,$
is independent of the temperature gradient ${\bf g}$ and determines the entropy,  the Cauchy stress, and the polarization vector through the relations
\begin{equation}   \label{eq:constRestreta00}
\overline{\eta}({\bf F}, \, \theta, {\bf E}^M)=
	-\partial_{\theta} \overline{\psi}({\bf F}, \, \theta, {\bf E}^M)  \, , 
\end{equation}

\begin{equation}   \label{eq:constRestrK00}
\mbox{\boldmath$\overline{\tau}$}({\bf F}, \, \theta, {\bf E}^M)=
\rho {\bf F} \partial_{{\bf F}} \overline{\psi}({\bf F}, \, \theta, {\bf E}^M)
\, , 
\end{equation}

\begin{equation}   \label{eq:constRestrP00}
\overline{\mbox{\boldmath$\mbox{\boldmath$\pi$}$}}({\bf F}, \, \theta, {\bf E}^M)
 =-\partial_{{\bf E}^M} \overline{\psi}({\bf F}, \, \theta, {\bf E}^M)\, .
\end{equation}

(ii) the Fourier inequality
\begin{equation}    \label{eq:RedDissIneq00}
{\bf q} \cdot {\bf g} \, \leq \, 0 \, 
\end{equation}
is satisfied.  \end{theorem}

A consequence of the reduced dissipation inequality (\ref{eq:RedDissIneq}) is that, just as in thermoelasticity, the {\it static heat flux} vanishes:
\begin{theorem}
The heat flux $\, {\bf q}\,$ vanishes for all termal equilibrium states 
$\,( {\bf F}, \, \theta, {\bf E}^M, \, {\bf 0} ) \in {\cal D}$, that is, 
\begin{equation}
\overline{{\bf q}}( {\bf F}, \, \theta, {\bf E}^M, \, {\bf 0} )
\,=\, {\bf 0}  \, .
\end{equation}
\end{theorem}
\subsection{Use of Invariant response functions (usual theory)}
The present section rewrites as \ref{subsection:Use of Invariant response functions} by dropping in the latter each $\dot {\bf q}$ where it appears as variable within a response function and $\dot {\bf q}$.
Hence we have the following
\begin{theorem}   \label{theorem:dissPrSpatialHAT00}
The Dissipation Principle is satisfied if and only if the following conditions hold:

(i) the objective free energy response function 
$\,\tilde \psi( {\bf E}, \, \theta, {\bf W},\, {\bf G} )  \,$
is independent of the temperature gradient ${\bf G}$ and determines the entropy,  the Cauchy stress tensor, and the polarization vector per unit mass through the relations (\ref{eq:TPsi})-(\ref{eq:piPsi});

(ii) the Fourier inequality (\ref{eq:RedDissIneq00})
is satisfied.  
  \end{theorem}

Again, Eqs. (\ref{eq:TPsi}), (\ref{eq:eulPolarVectorUnVolM}) and (\ref{eq:piPsi}) yield the Cauchy stress expression  
(\ref{eq:TPsiExpr})
and for its antisymmetric portion is (\ref{eq:TPsiExprSKW}),
that coincides with (3.24) of \cite{RefTier1}.
\subsection{Internal Dissipation and Entropy Equality (usual theory)}   
In Section \ref{section:Internal Dissipation and Entropy Equality}
just the same proofs of Theorem \ref{theorem:diss} and Theorem \ref{theorem:LocRev}
with no change yield the proofs of the theorems below.
\begin{theorem}\label{theorem:diss00}
Along any local process of $B$ we have
\begin{equation}\label{eq:Thdiss00}
	\delta_o= 0  \, .
\end{equation}
\end{theorem}
\begin{theorem}\label{theorem:LocRev00}
Along any local process of $B$ the following entropy equality
\begin{equation}\label{eq:EntEq00}
	\rho \dot \eta =  \rho \frac{r}{\theta} - \frac{div{\bf q}}{\theta} \, .
\end{equation}
holds.
\end{theorem}


\end{document}